\documentclass[amsmath,amssymb,superscriptaddress,reprint]{revtex4}
\usepackage{graphicx}

\usepackage{dcolumn}
\usepackage{bm}
\usepackage{aas_macros}
\usepackage{times}
\usepackage{longtable}
\usepackage{siunitx}
\usepackage{booktabs}
\usepackage{etoolbox}
\usepackage{lipsum}
\usepackage{ulem}
\usepackage[hidelinks]{hyperref}
\robustify\itshape
\usepackage[dvipsnames]{xcolor}
\usepackage{xargs}
\usepackage[colorinlistoftodos]{todonotes}  
\newcommandx{\unsure}[2][1=]{\todo[linecolor=red,backgroundcolor=red!25,bordercolor=red,#1]{#2}}
\newcommandx{\change}[2][1=]{\todo[linecolor=blue,backgroundcolor=blue!25,bordercolor=blue,#1]{#2}}
\newcommandx{\info}[2][1=]{\todo[inline,linecolor=OliveGreen,backgroundcolor=OliveGreen!25,bordercolor=green,#1]{#2}}
\newcommandx{\improvement}[2][1=]{\todo[linecolor=Plum,backgroundcolor=Plum!25,bordercolor=Plum,#1]{#2}}

\newcolumntype{.}{D{.}{.}{1}}
\newcolumntype{X}{D{X}{X}{1}}

\begin{document}

\newcommand{\reaction}[6]{\nuc{#1}{#2}(#3,#4)\/\nuc{#5}{#6}}
\newcommand{\nuc}[2]{\ensuremath{^{#1}}#2}
\newcommand{\fpg}[0]{\reaction{18}{F}{p}{$\gamma$}{19}{Ne} }
\newcommand{\fpa}[0]{\reaction{18}{F}{p}{$\alpha$}{15}{O} }
\newcommand{\nehe}[0]{\reaction{20}{Ne}{$^{3}$He}{$\alpha$}{19}{Ne} }
\newcommand{\Ne}{$^{19}$Ne }
\newcommand{\Jpi}{J$^{\pi}$ } 
\newcommand{\Deg}{$\,^{\circ}$ }
\newcommand{\Fpa}{$^{18}$F(p, $\alpha$)$^{15}$O }
\newcommand{\reac}{$^{20}$Ne($^3$He, $^4$He)$^{19}$Ne }
\newcommand{\fluor}{$^{18}$F }
\newcommand{\mirror}{$^{19}$F }
\newcommand{\fpp}{$^{18}$F+p}

\newcommand{\Erlab}[1]{\ensuremath{E_{r}^{\text{lab}} = #1}~keV}
\newcommand{\Ercm}[1]{\ensuremath{E_{r}^{\text{c.m.}} = #1}~keV}
\newcommand{\Ercmb}[1]{\ensuremath{\mathbf{E_{r}^{\textbf{c.m.}} = #1}}~\textbf{keV}}
\newcommand{\Ex}[1]{\ensuremath{E_{x} = #1}~MeV}
\newcommand{\Exb}[1]{\ensuremath{\mathbf{E_{x} = #1}}~\textbf{keV}}
\renewcommand{\pg}[0]{\reaction{39}{K}{p}{$\gamma$}{40}{Ca} }
\newcommand{\pa}[0]{\reaction{39}{K}{p}{$\alpha$}{36}{Ar} }
\newcommand{\Jpib}[2]{\ensuremath{\mathbf{J^{\pi} = #1^{#2}}} }
\newcommand{\res}{$^{18}$F + p }
\def\sun{\odot}

\title{Spin-parities of sub-threshold resonances in the \Fpa reaction}

\author{F. Portillo} 
\author{R.~Longland}
\email[]{richard\_longland@ncsu.edu}
\affiliation{Department of Physics, North
  Carolina State University, Raleigh, NC 27695, USA}
\affiliation{Triangle Universities Nuclear Laboratory, Durham, NC
  27708, USA}
\author{A.~L.~Cooper}
\altaffiliation{Present Address: Los Alamos National Laboratory, Los Alamos, NM 87545, USA}
\author{S.~Hunt}
\affiliation{Department of Physics and Astronomy, University of North
  Carolina at Chapel Hill, Chapel Hill, NC 27599, USA}
\affiliation{Triangle Universities Nuclear Laboratory, Durham, NC
  27708, USA}
\author{A.~M.~Laird}
\affiliation{Department of Physics, University of York, York YO10 5DD, UK}
\author{C.~Marshall}
\altaffiliation{Present address: Institute of Nuclear \& Particle Physics, Department of
  Physics \& Astronomy, Ohio University, Athens, Ohio 45701, USA}
\author{K.~Setoodehnia}
\altaffiliation{Present address: Facility for Rare Isotope Beams, Michigan State University, East Lansing, MI 48824, USA}
\affiliation{Department of Physics, North
  Carolina State University, Raleigh, NC 27695, USA}
\affiliation{Triangle Universities Nuclear Laboratory, Durham, NC
  27708, USA}

\begin{abstract}
The \Fpa reaction is key to determining the $^{18}$F abundance in classical novae. However, the cross section for this reaction has large uncertainties at low energies largely caused by interference effects. Here, we resolve a longstanding issue with unknown spin-parities of sub-threshold states in \nuc{19}{Ne} that reduces these uncertainties. The \reac neutron pick-up reaction was used to populate \Ne excited states, focusing on the energy region of astrophysical interest ($\approx$ 6 - 7 MeV). The experiment was performed at the Triangle Universities Nuclear Laboratory using the high resolution Enge split-pole magnetic spectrograph. Spins and parities were found for states in the astrophysical energy range. In particular, the state at 6.133 MeV (\Ercm{-278}) was found to have spin and parity of $3/2^+$ and we confirm the existence of an unresolved doublet close to 6.288 MeV (\Ercm{-120}) with \Jpi = $1/2^+$ and a high-spin state. Using these results, we demonstrate a significant factor of two decrease in the reaction rate uncertainties at nova temperatures.
\end{abstract} 


\maketitle

\section{Introduction}
\label{intro}

Classical novae are the observational signature of thermonuclear explosions that occur in binary systems consisting of a white dwarf and typically a Main Sequence star (see Ref.~\cite{hernanz2005} for an overview). When $\approx$ 10$^{-5}$\ M$_{\odot}$ of hydrogen-rich material is accreted by the white dwarf at a low rate (e.g., 10$^{-11}$\ M$_{\odot}$yr$^{-1}$), the base of the envelope reaches the conditions necessary to generate a thermonuclear explosion, where the temperature rises dramatically from 0.1 - 0.4 GK a thermonuclear runaway occurs. This explosion releases an energy of $\approx$ 10$^{45}$ ergs and material containing elements up to mass $A = 40$ \cite{bode}. The nucleosynthesis path proceeds close to the valley of stability following break-out from the carbon-nitrogen-oxygen (CNO) cycle in a series of (p,$\gamma$) and (p,$\alpha$) reactions, and $\beta^+$ decays.

Among the elements synthesized in classical novae, the radioactive \fluor ($\tau_{1/2} = $110 min) decays by $\beta^+$ decay, thus emitting 511 keV $\gamma$ rays. This is of particular interest because the decay is expected to occur shortly after the explosion when the remnant becomes transparent to gamma rays. However, to date no $\gamma$ rays have been observed at these energies, in part due to the low sensitivity of the instruments used for their detection (e.g, INTEGRAL \cite{integral-hernanz}) and in part because these $\gamma$ rays are emitted before the nova visual peak \cite{novaquote}. Detection of these low energy $\gamma$ rays would help improve nova theoretical models, but will be an unreliable constraint unless uncertainty in the nuclear reactions producing fluorine is significantly reduced. The \Fpa reaction rate is key among these.

The source of the uncertainty in the \Fpa reaction is mostly due to interference effects between low energy resonances close to the proton threshold and higher energy resonances. Many of the resonances in the \fpa and \fpg reactions were first identified and compared with the mirror \mirror in 1998 by Utku \textit{et al.}~\cite{utku}. Following that, a series of measurements have further constrained the reaction rates through direct cross section measurements (Refs. \cite{bardayan2001,bardayan2002,desereville,beer}), differential cross sections~\cite{murphy,Mountford2012}, and particle transfer or charge exchange reactions~\cite{adekola,laird,Parikh2015,bardayan2017,kahl2019,riley}, including some coincident $\gamma$-ray measurements~\cite{hall2019,hall2020}. These results have been thoroughly compiled previously by Nesaraja \textit{et al.} 2007~\cite{nesaraja} and most recently by Kahl \textit{et al.} 2021~\cite{kahl2021}. However, despite these efforts, the reaction rate of the \Fpa reaction contains large uncertainties.


In a recent result, Ref.~\cite{kahl2019} found that the \Jpi quantum numbers of the state below the \Ne proton threshold at 6.133 MeV (\Ercm{-278} resonance) should be either $1/2^+$ or $3/2^+$, confirming the presence of an s-wave resonance close to and below the proton threshold. The other \Ne excited state, close to the proton-threshold, predicted to correspond to an s-wave resonance, has been proposed to exist around the 6.29 MeV energy region (\Ercm{-120} resonance). The energy and the spin and parity of this state have been much debated, where assignments of low spin \cite{bardayan2015} or high spin (e.g., \cite{laird}, \cite{hall2019}, \cite{riley}) have been proposed. Parikh et al. \cite{Parikh2015} suggested that this state could be either a doublet or a broad state. These ambiguities lead to large uncertainties arising from strongly different interference effects in the astrophysical energy region, as explored by the study of Kahl \textit{et al.} 2021~\cite{kahl2021}.

In this work, we present results that significantly constrain the interference possibilities by clarifying the spin-parities of the sub-threshold resonances at \Ercm{-278} and \Ercm{-121}. These were obtained using the neutron pick up reaction \nehe to populate \Ne excited states in order to determine their energies and spins and parities. In Section \ref{methods} experimental details of this work are described. Section \ref{sec:analysis} describes our analysis, for which the results for \Ne states are discussed in section \ref{results-discussion}. The impact of these findings on the cross section and their impact on the \Fpa reaction rates at classical nova temperatures is presented in Sec. \ref{sec:r-matrix-analysis}. A summary and conclusions are given in Section \ref{sec:conclusions}.

\section{Experiment Details}
\label{methods}

The implanted $^{20}$Ne transmission targets used in the experiment were fabricated using an electron cyclotron resonance ion source \cite{andrew-ecr} to implant $^{20}$Ne at an energy of 25 keV directly into a 40 $\pm 4$ $\mu$g/cm$^2$ thick carbon foil. The foil was mounted on a rectangular glass slide (3.7 cm x 2.5 cm) from which a total of four targets were obtained.
\begin{figure}
\includegraphics[width=0.5\textwidth]{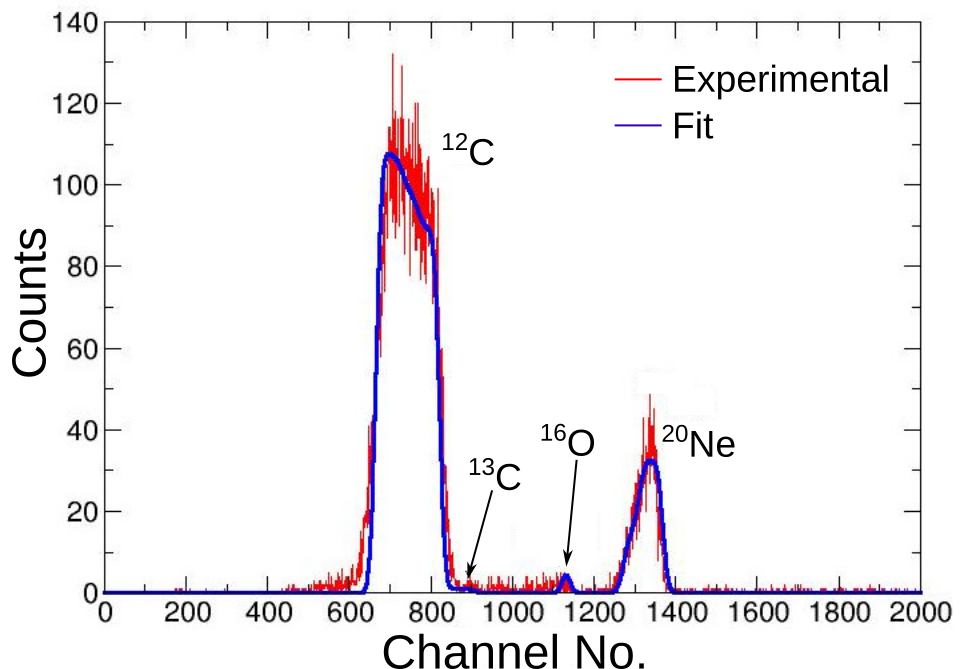}
\caption{RBS spectrum. The red and blue lines are the experimental data and SIMNRA~\cite{SIMNRA} simulation, respectively. The analysis done shows that, apart from the natural carbon substrate, the target contains a 3.6 $\pm$ 0.4 $\mu$g/cm$^2$ thick $^{20}$Ne layer and an $^{16}$O contamination layer 0.22 $\pm$ 0.02 $\mu$g/cm$^2$ thick located at the front.  }
\label{rbs}
\end{figure}
To characterize the targets, we used the Rutherford Backscattering Spectrometry technique (RBS) on one of the targets by accelerating a 2 MeV $^4$He$^{++}$ beam and placing a silicon surface barrier detector at 165\Deg, with respect to the incident beam, to detect the back-scattered $\alpha$-particles. Figure \ref{rbs} shows the RBS spectrum for one of the $^{20}$Ne targets. Based on analysis using the software package \texttt{SIMNRA} \cite{SIMNRA}, the targets were determined to consist of carbon, with a 3.6 $\pm$ 0.4 $\mu$g/cm$^2$ thick region of $^{20}$Ne, located close to the center, a relatively thin 0.22 $\pm$ 0.02 $\mu$g/cm$^2$ thick $^{16}$O contamination layer located at the front. The $^{16}$O contamination was also present on unimplanted carbon foils, indicating that it has a common source.

The neutron pick up reaction $^{20}$Ne($^3$He,$^4$He)$^{19}$Ne at a beam energy of 21 MeV was used to populate excited states in $^{19}$Ne. This beam energy was chosen to ensure the direct nature of the pick-up reaction. The Q-value of this reaction is 3712.32 (16) keV~\cite{AME2016}. The experiment was performed at the Triangle Universities Nuclear Laboratory (TUNL) by using its 10 MV FN Tandem Van de Graaff accelerator to accelerate a $^3$He$^{++}$ beam to an energy of 21 MeV with a mean intensity on target of 600 pnA. The dead-time registered while collecting data was less than 1\% at a rate of $\approx$ 200 Hz. 
Before impinging on target, the beam was tuned through a 1-mm-diameter collimator. Data were collected for both implanted $^{20}$Ne and a natural carbon target used for background reference. Reaction products entered the TUNL Enge split-pole magnetic spectrograph \cite{Kiana_Enge} whose magnetic field was set at B = 0.81 T to focus reaction products into the detector's focal plane. The spectrograph was set at six different laboratory angles ($\theta_{lab} $= 9\Deg, 12\Deg, 20\Deg, 22\Deg, 25\Deg, and 27\Deg) and its entrance aperture subtended a solid angle with respect to the target of 1 msr for all angles. The beam was on target for about 10 hours for each angle. Target degradation was monitored hourly using the beam current normalized intensity of the peak corresponding to the strongly populated 6.0143 (11) MeV excited state. Degradation was corrected for during analysis.

\begin{figure*}
\label{2dspectrum}
    \includegraphics[width=0.9\textwidth]{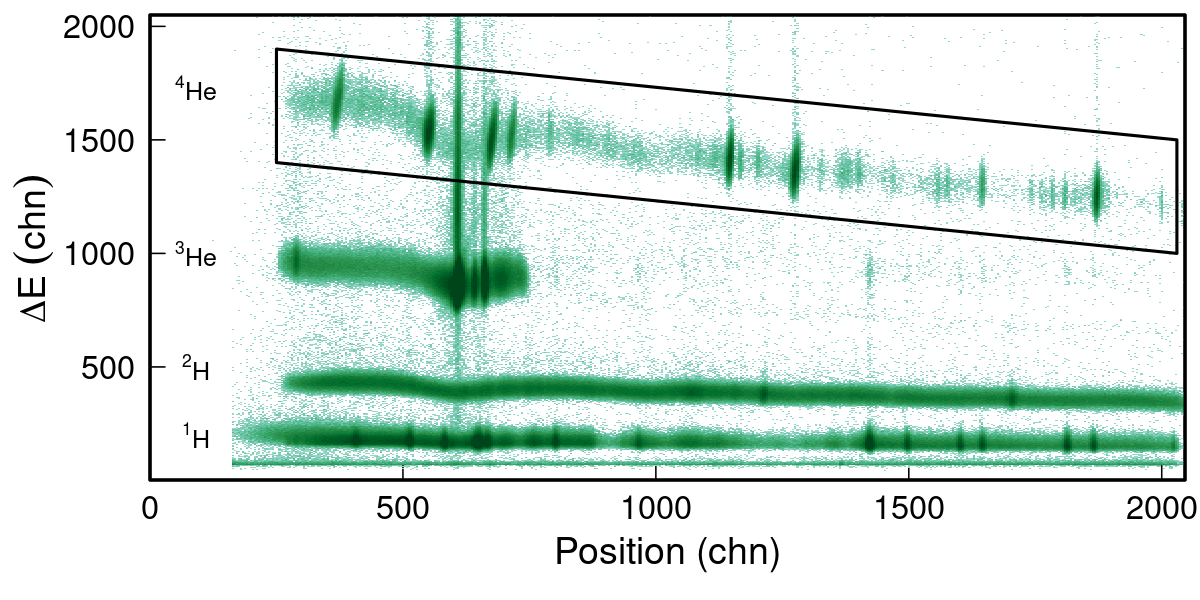}
    \caption{\label{2dspec} Two-Dimensional energy loss vs front position spectrum collected at $\theta_{Lab}=$ 20$^{\circ}$ from signals coming from these sections of the focal plane detector. Bands of events coming from protons, deuterons, $^3$He, and $^4$He are indicated, with the software gate used for $\alpha$-particles shown as a black polygon. }
\end{figure*}

A position-sensitive focal plane detector was used to detect the outgoing reaction products that entered the spectrograph. This detector consists of two position sensitive sections, one energy loss section, and a total energy section. The position sensitive sections are avalanche counters consisting of five equally spaced high voltage wires placed in the isobutane gas filled region across the detector. Their position resolution corresponds to an energy resolution of about 20 keV for the present experiment. These two sections are separated by an energy loss section, which is a proportional gas counter consisting of a single high voltage wire also within an isobutane gas filled region. Finally, the total energy section is a plastic scintillator attached to a photomultiplier tube and is used to register the particles' residual energy. Details of the focal plane detector can be found in \cite{caleb-paper}.

\section{Analysis}
\label{sec:analysis}

\begin{figure*}
  \includegraphics[width=0.98\textwidth]{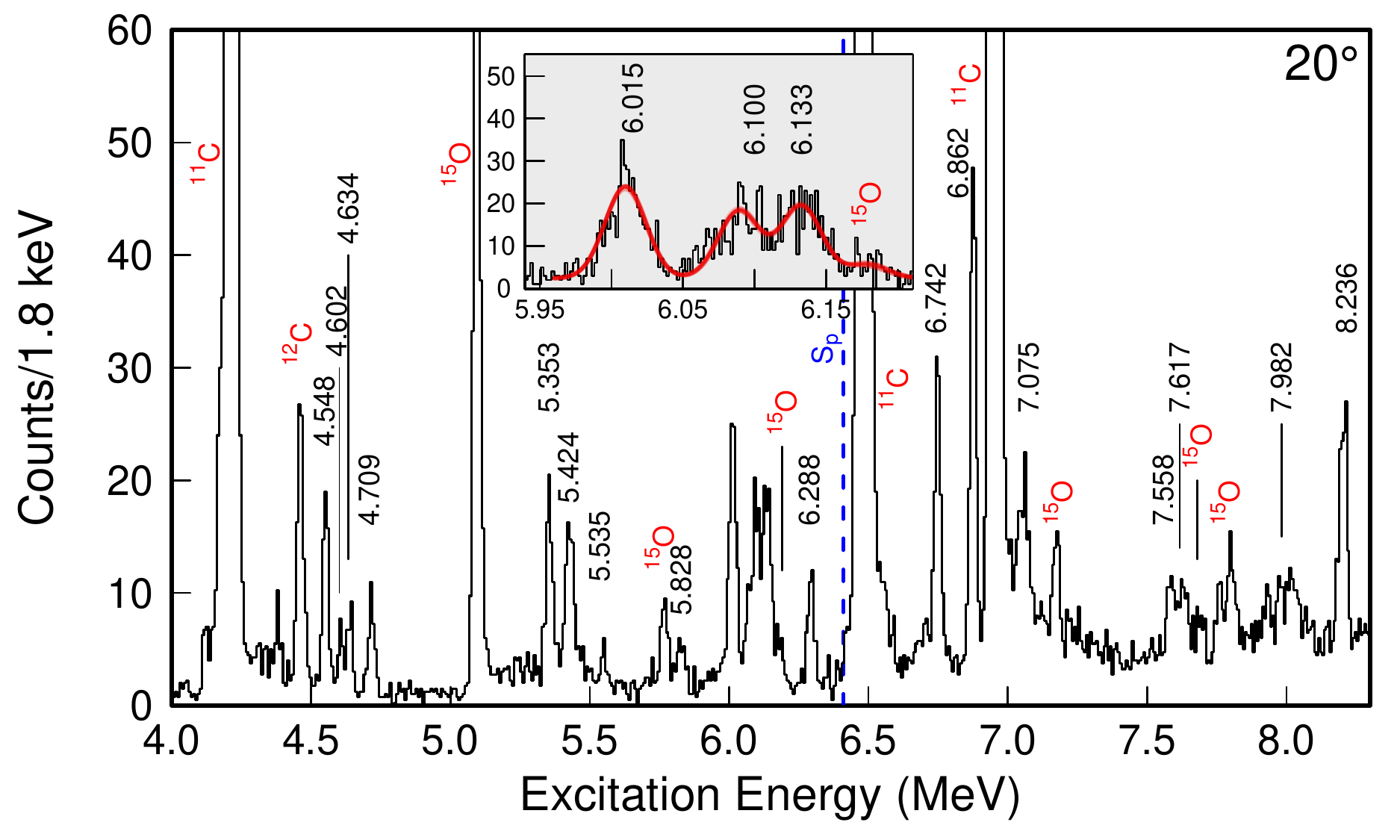}
  \caption{\label{spectrum} 
    Energy calibrated $\alpha$-particle gated spectrum at $\theta_{\text{lab}} = 20^{\circ}$. The  peaks corresponding to $^{19}$Ne excited states are labeled with the energies (in MeV) in column 2 of Tab. \ref{table1}. The peaks with a red label correspond to excited states coming from reactions involving the substrate or contaminants. The inset figure highlights the region around the sub-threshold resonances, which strongly influence the interference effects at high energies. The vertical dashed blue line denotes the proton separation energy.} 
\end{figure*}

Alpha particles were selected by setting a software gate on the energy loss vs front position spectrum. An example gathered at $\theta_{Lab}=$20$^{\circ}$ is shown in Fig.~\ref{2dspec}. By setting an off-line software gate to data in the $^4$He band we generated the 1-dimensional spectrum  shown in Fig.~\ref{spectrum}. Peak positions and intensities for the $\alpha$-particle spectra at each angle were obtained by fitting Gaussian peaks with a linear function to describe the local background. In areas with many peaks, several Gaussian peaks were fitted with a common width to help constrain the fit. 

To determine the energies of the observed \Ne excited states, the code \texttt{SPANC} \cite{spanc} was employed to perform an internal calibration from known calibration states in \Ne. A second degree polynomial model was used to fit particle position on the focal plane as a function of the centroid channel of calibration peaks. To account for energy loss in the target, SPANC includes angle-dependent entrance and exit particle energy loss estimation. From the polynomial fits, excitation energies at each spectrograph measurement angle could be extracted. Our reported excitation energies in the third column of Tab.~\ref{table1} were then determined through a weighted average over all angles. Due to systematic effects, the uncertainty in the weighted average is constrained to be no smaller than the smallest single-angle energy uncertainty. To account for an estimated detector response uncertainty, the weighted average energies are summed in quadrature with an additional 2 keV uncertainty. 

The spin and parities of states analyzed in this work were constrained by comparing the experimental angular distribution for each state with theoretical angular distributions. 
Theoretical angular distributions were computed assuming that the pick up reaction was a one-step transfer reaction. We used the Distorted Wave Born Approximation (DWBA) finite-range code FRESCO \cite{FRESCO} with the global optical potential parameters shown in Tab.~\ref{optical-table}. The depth of the potentials used for the neutron binding was varied to reproduce the neutron binding energy. 

\begin{table*}
\centering
\caption{\label{optical-table} Global optical potential parameters used in the DWBA analysis.}
\resizebox{\textwidth}{!}
{\begin{tabular}{  c c c c c c c c c c c c c c c}
   \toprule\\
   Channel                & r$_C$(fm) & V$_R$ (MeV) & r$_R$ (fm) & a$_R$ (fm) & V$_I$ (MeV) & r$_I$ (fm) & a$_I$ (fm) & V$_D$ (MeV) & r$_D$ (fm) & a$_D$ (fm) & V$_{so}$ (MeV) & r$_{so}$ (fm) & a$_{so}$ (fm) \\
   \hline                                                                                                                                                                                                      \\
   $^{3}$He+$^{20}$Ne$^b$ & 1.3       & 148.33      & 1.2        & 0.72       & 34.77       & 1.4        & 0.88       & -           & -          & -          & 2.5            & 1.2           & 0.72          \\
   $^{4}$He+$^{19}$Ne$^c$ & 1.35      & 160.12      & 1.34       & 0.66       & -           & 1.43       & 0.56       & 23.38       & 1.29       & 0.64       & -              & 1.27          & 0.85          \\
   $^3$He + $^{19}$Ne$^b$ & 1.3       & 160.12      & 1.34       & 0.66       & 32.45       & 1.43       & 0.56       & -           & -          & -          & -              & -             & -             \\
   $^{19}$Ne+n            & 1.2       & $^a$        & 1.12       & 0.79       & -           & -          & -          & -           & -          & -          & 3.0            & 1.27          & 0.75          \\
   $^{3}$He+n             & 2         & $^a$        & 1.17       & 0.69       & -           & -          & -          & -           & -          & -          & 6.0            & 1.17          & 0.69          \\
   \hline                                                                                                                                                                          \\
 \end{tabular}
}
\footnotetext{Adjusted to produce binding energy}
\footnotetext{From Perey and Perey 1976~\cite{perey}}
\footnotetext{From Su and Han 2015~\cite{xinglobal}}
\end{table*}

\section{Results and Discussion}
\label{results-discussion}

\begin{table}
\label{table}
\centering
\caption{\Ne excited states energies (in MeV) observed in this work. The columns show, from left to right, the energies reported by NNDC \cite{NNDC} compilation; our compiled energies since the NNDC values were published (1997-2020), i.e., Refs. \cite{utku}, \cite{tan2009}, \cite{dalouzy}, \cite{murphy}, \cite{laird}, \cite{bardayan2015}, \cite{bardayan2017}, \cite{kahl2019}, \cite{hall2019}, and\cite{kahl2021}; and the energies determined in this work (see text for details). The compiled energies marked with an asterisk (*) and in \textit{italics} were used for calibrations in this work.} 
\begin{tabular}{ l c c }
\toprule \hline \\
NNDC        &   Compilation    &      Present Work  \\
\hline\\                                                                       
 1.5360 (4)  &    1.535 (3)      &      1.532 (3)   \\
 1.6156 (5)  &    1.6156 (4)     &      1.613 (3)   \\
 2.7947 (6)  &   *\textit{2.7944 (5)}  &    -       \\
 4.0329 (24) &    4.0344 (6)     &      4.0347 (23) \\
 4.140  (4)  &    4.1431 (7)     &      4.1421 (22) \\
 4.1971 (24) &    *\textit{4.1998 (7)} &    -       \\
 4.549  (4)  &    *\textit{4.5477 (9)} &    -       \\
 4.600  (4)  &    4.6021 (6)     &      4.6017 (25) \\
 4.635  (4)  &    4.6341 (9)     &      4.633 (3)   \\
 4.712  (10) &   *\textit{4.7088 (17)} &    -       \\
 5.092  (6)  &    5.0906 (23)    &      5.0908 (25) \\
 5.351  (10) &    5.353 (5)      &      5.351 (3)   \\
 5.424  (7)  &    5.424 (5)      &      5.424 (3)   \\
 5.539  (9)  &    5.535 (7)      &      5.543 (3)   \\
 5.832  (9)  &    *\textit{5.828 (7)}  &    -       \\
 6.013  (7)  &   *\textit{6.0147 (15)} &    -       \\
 6.092  (8)  &    6.0998 (9)     &      6.092 (3)   \\
 6.149  (20) &    6.1327 (23)    &      6.133 (3)   \\
  6.288  (7)  &    6.2883 (14)    &      6.284 (3)   \\
  6.699 (3)  &    6.699 (3)      &       Unobserved          \\
 6.742  (7)  &    6.7420 (12)    &      6.738 (3)   \\
 6.861  (7)  &    6.8619 (15)    &      6.861 (4)   \\
 7.067  (9)  &    *\textit{7.0747 (9)} &   -        \\
 7.531  (15) &    7.558 (3)      &      7.549 (5)   \\
 7.616  (16) &    7.617 (3)      &      7.605 (5)   \\
 7.700  (10) &    *\textit{7.706 (17)} &   -        \\
 7.788  (10) &    *\textit{7.788 (9)}  &   -        \\
 7.994  (15) &    7.982 (6)      &      7.970 (4)   \\
 8.069  (12) &    8.069 (9)      &      8.057 (5)   \\
 8.236  (10) &   *\textit{8.236 (10)}  &   -        \\
\hline \hline\\                       
\end{tabular}
\label{table1}
\end{table}

The energies are reported in Tab.~\ref{table1}. For ease of comparison, we first list the energies as compiled in the National Nuclear Data Center (NNDC) \cite{NNDC}. The second column corresponds to an average of the NNDC values and all following energies reported in the literature from the years 1995 to 2021 (\cite{utku},  \cite{kozub}, \cite{chae}, \cite{nesaraja}, \cite{dalouzy}, \cite{murphy}, \cite{tan2009}, \cite{laird}, \cite{bardayan2015}, \cite{bardayan2017}, \cite{hall2019}, \cite{kahl2019}, \cite{kahl2021}). The third column contains energies obtained in this work using the procedure described above. 


Each of the observed levels is discussed individually below for the energy region relevant for classical novae (6 - 7 MeV). A tentative analog state assignment in $^{19}$F is made for most of the observed \Ne excited states by assuming isospin symmetry and by considering their energies. This follows the same procedure as other recent works (Refs.~\cite{bardayan2015,hall2020,kahl2021}, for example). Mirror pairs were assigned below 6 MeV by systematically compiling the levels in each mirror nuclei comparing the \Jpi values of the analog pair states. In compiling mirror assignments, we find an average mirror energy difference (MED) of 70 $\pm$ 50 keV between the pairs, which we use as a guide in our discussion below. However, we should stress that the information is not complete. There could be intruder states, or missing states from the mirror, \mirror, that may complicate the picture, and some well-known states still do not have an assigned mirror partner (see below). Future work should concentrate on a full accounting of mirror states between \nuc{19}{Ne} and \nuc{19}{F}, and their applicability to calculating the \Fpa cross section.


\noindent \textbf{6.0147 (15) MeV, \Ercmb{-395}, \Jpi = (1/2$^-$, 3/2$^-$):}
\label{6014}
This state was suggested by Ref.~\cite{riley} to be an unresolved doublet consisting in two states approximately around 6.008 MeV and a 6.014 MeV. However, the 6.008 MeV state was proposed to be very broad ($\Gamma = 124 (25)$~keV). We were not able to resolve this lower-energy state, and furthermore observed no broadening of the peak in our spectra. The angular distribution obtained for this state indicates that only a single state was strongly populated. The angular momentum transfer is clearly described by $\ell = 1$, so the spin and parity of this state could be either \Jpi = $1/2^-$ or \Jpi = $3/2^-$. The average MED suggests that the analog state in \mirror could be the 6.088 MeV level, which has \Jpi = $3/2^-$. There is no mirror state with a \Jpi = $1/2^-$ within the average MED (the closest is at 6.429 MeV with an MED=414 keV), but we nevertheless maintain a tentative \Jpi = $(1/2^-, 3/2^-)$. 

\begin{figure*}
  \includegraphics[width=0.7\textwidth]{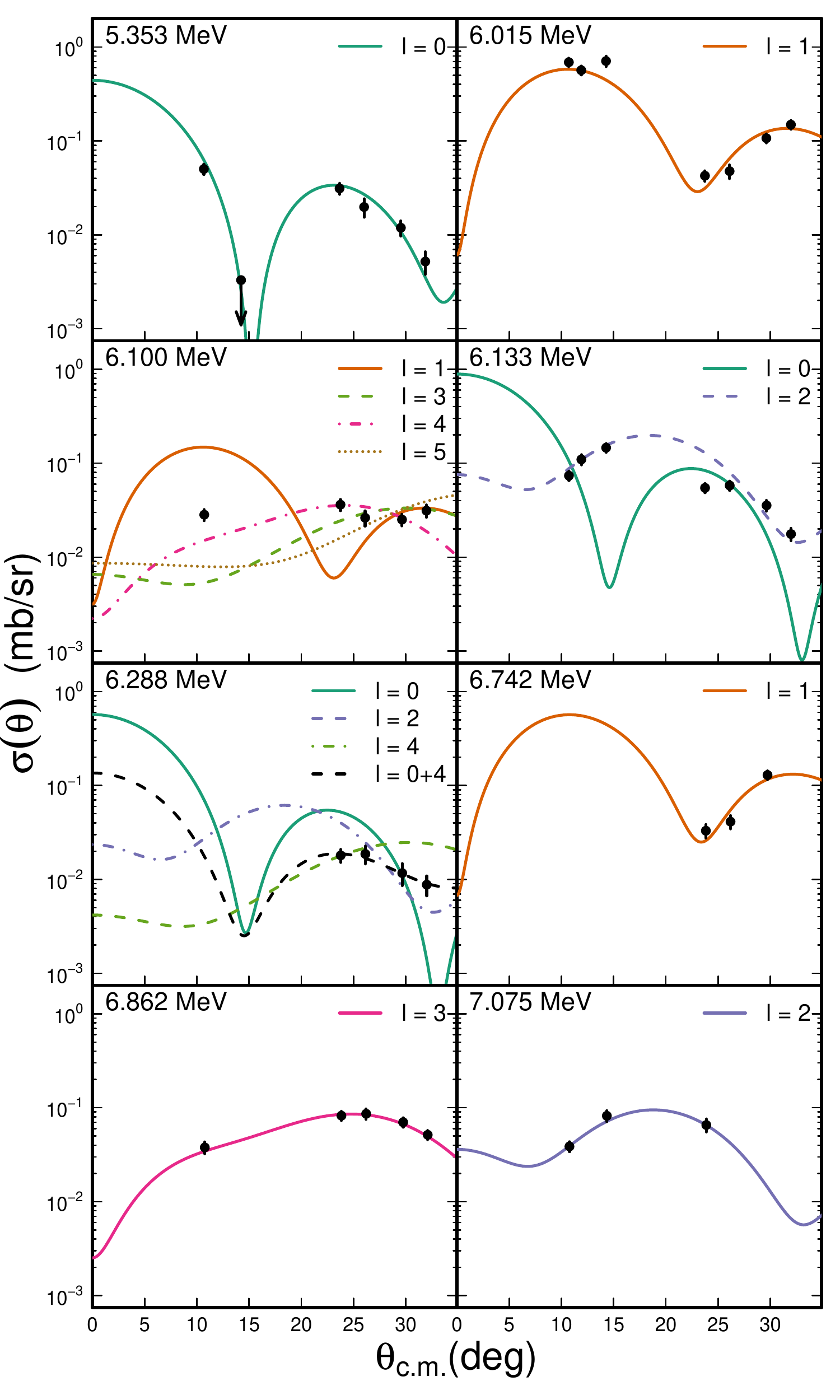}
 \caption{{Angular distributions obtained from the \reac} experiment for \Ne excited states with energies of astrophysical interest (6 - 8 MeV). The curves were calculated with \texttt{FRESCO}. Each panel shows the \Ne level energy together with the angular momentum transfers that were found to best describe the data. The downward-facing arrow for the state at 5.351 MeV indicates an upper limit.}
 \label{ang_dist}
\end{figure*}

\noindent \textbf{6.0998 (9) MeV, \Ercmb{-310}, \Jpi = (5/2$^-$, 7/2$^-$,7/2$^+$, 9/2$^+$):}
\label{6090}
The peak corresponding to this state appeared as a member of a resolved doublet with the 6.133 MeV state. Fig. \ref{ang_dist} shows the experimental angular distribution for this state together with the theoretical curves for the possible $\ell$-values that correspond to the \Jpi values of close energy analog states within the average MED, namely, 6.070 ($7/2+$), 6.088 ($3/2-$), 6.100 ($9/2-$), and 6.161 ($7/2-$) MeV. The angular distribution for this state is very flat, suggesting a high-spin state. The experimental data are best described  by the $\ell=3$ curve and to a lesser extent to the one corresponding to $\ell=4$. This is in slight tension with the results of Laird \textit{et al.} \cite{laird}. 
 
\noindent \textbf{6.1327 (23) MeV, \Ercmb{-277}, \Jpi = 3/2$^+$:}
\label{6132-state}
This state corresponds to the \fpp\ sub-threshold resonance at \Ercm{-277}. The spin-parity of this state has long been discussed, and was recently narrowed down by Kahl \textit{et al.} \cite{kahl2019}. In their work they populated \Ne states resulting from Gamow-Teller transitions in the \mirror($^3$He,t)\Ne reaction. They observed a state at 6.133 MeV that must have \Jpi = $1/2^+$ or $3/2^+$. As will be discussed in Section \ref{sec:r-matrix-analysis}, the spin and parity of this state impacts the \Fpa reaction cross section uncertainty at low energies due to interference effects.

This state was observed in our work as one of the members of a resolved doublet whose other member is the state at 6.100 MeV discussed above. Fig.~\ref{ang_dist} presents the experimental angular distribution for this state together with curves corresponding to the two angular momentum transfer possibilities established by Ref.~\cite{kahl2019} (i.e., $\ell=0$ and $\ell=2$). The angles between $\theta_{c.m.}$ = 11$^{\circ}$ - 18$^{\circ}$ and $\theta_{c.m.}$ = 30$^{\circ}$ - 36$^{\circ}$ are where the two theoretical angular distributions differ appreciably (by more than one order of magnitude in the cross section). Consequently, data at $\theta_{c.m.}\approx$ = 14$^{\circ}$ and  $\theta_{c.m.}\approx$ = 32$^{\circ}$ are key to determining that $\ell=2$. This is in clear contrast to the state at 5.353 MeV corresponding to $\ell=0$, where only an upper limit could be obtained at $\theta_{c.m.}=14^{\circ}$. In combination with Ref.~\cite{kahl2019}, we can confirm that this state has a spin-parity of \Jpi = $3/2^+$. We could not, however, locate a mirror level in \mirror within the calculated average MED $70 \pm 50$ keV, corresponding to a state in \nuc{19}{F} at about 6.2 MeV. The closest are either at 5.501 MeV (MED=630 keV) or the pair at 6.497 and 6.527 MeV (MED=365 and 394 keV, respectively). This finding suggests that there may be an undiscovered \Jpi{3/2}{+} state in the mirror nucleus. A full re-accounting for mirror states should be undertaken to assign mirror pairs, but this is outside the scope of the present work. 

\noindent \textbf{E$_x$= 6.288 MeV (\Ercmb{-122}) region:}
\label{6290-state}
A state in this region corresponds to the \Ercm{-122} sub-threshold resonance closest to the proton threshold. The spin-parity of this state has been discussed extensively owing to its impact on interference patterns in the cross section. Refs. \cite{Parikh2015} and~\cite{hall2020} suggested that there is likely an unresolved doublet of high and low spin states at this energy.

The measured angular distribution for a peak corresponding to this energy region is shown in Fig.~\ref{ang_dist}. Unfortunately owing to the presence of contaminant peaks in the spectrum, data were only extracted at higher angles for this state. They are not described by a single transferred angular momentum ($\ell=0$, $\ell=2$, or $\ell=4$), but rather support the hypothesis that there is a doublet. Since the expected separation of $\approx 10$~keV is far below the resolution of our detector, we do not see any obvious broadening of the peak. However, the data are best described by the combination of $\ell=0+4$ in the proportion 0.40 (10) + 0.60 (10), although all angular momenta with $\ell \ge 4$ for the second member are supported by our data. This supports the results of Refs. \cite{Parikh2015} and~\cite{hall2020} that a closely-spaced doublet is present at this energy. The latter of these results placed a likely value of \Jpi = $11/2^+$ from observed $\gamma$-ray decays. However, the nearest mirror state in \nuc{19}{F} is at 6.500 MeV with MED=212 keV, placing it in tension with other mirror assignments. We therefore adopt an assignment of \Jpi$=7/2^+$ in our cross section calculations to demonstrate the maximum effect of this level. 
The \Jpi = $1/2^+$ component of this doublet has the most impact on the reaction rates for the \Fpa reaction, since it interferes with the broad 1.4 MeV resonance \cite{bardayan2015}. The \Jpi $\ge$ $7/2^+$ component has negligible impact since it implies a proton orbital angular momentum of $l_p \ge 2$ i.e., a higher angular momentum barrier in the \Fpa reaction. If a spin-parity of \Jpi$ = 11/2^+$ were assumed, as suggested by Ref. \cite{hall2020}, the contribution would be even less.

Based on the calculated average MED and on the \Jpi values found, the only two analog states in \mirror that have these \Jpi quantum numbers and that fall within the 6.2 - 6.3 MeV energy region are the 6.255 MeV ($1/2^+$ with MED=33 keV) and 6.330 MeV ($7/2^+$ with MED=42 keV) levels.

\noindent \textbf{E$_x$= 6.416 -- 6.459 MeV (\Ercmb{6 - 49}):}
\label{theblob}
Structure corresponding to states in this region was only observed at 25 and 27 degrees. We were also unable to extract any parameters for these states corresponding to \Ercm{6 - 49} owing to their close proximity to each other. We adopt the following from a combination of Refs. \cite{kahl2021}, \cite{hall2020}, and references therein. The 6.416 MeV state was assigned a spin-parity of $3/2^+$ by Kahl \textit{et al.}~\cite{kahl2021}, which is in tension with the assignments of Refs. \cite{adekola,laird,hall2020}. We adopt the 6.439 MeV, $1/2^-$ state parameters from Ref.~\cite{bardayan2015}. Hall \textit{et al.} proposed two $3/2^+$ states at 6.423 MeV and 6.441 MeV~\cite{hall2020}. As noted by Ref.~\cite{kahl2021}, this number of $3/2^+$ states near the proton threshold exceeds the number of known $3/2^+$ states in the mirror \nuc{19}{F}. As with the 6.132 MeV state, this suggests that there may be undiscovered $3/2^+$ states in \nuc{19}{F}. We adopt the upper limit values assigned by Ref.~\cite{hall2020} for the 6.441 MeV state. Finally, we adopt the 6.459 MeV, $5/2^-$ state parameters from Ref.~\cite{laird}.

Attempts were made to fit multiple peaks to the peak structure in the region. By constraining the peak widths using neighboring states, one or more peaks were fit to the spectrum at 25 degrees, where the 6.4 MeV region is free of background contamination. No more than a single peak at 6.452 MeV is consistent with our data. It is possible that the selectivity of the ($^3$He,$^4$He) reaction only populates a single state, or perhaps the number of counts obtained for these states is not high enough to observe the other states in the region.

\noindent \textbf{E$_x$= 6.699 (3) MeV, \Ercmb{289}, \Jpi = $(5/2^+)$:}
\label{6699}
The state corresponding to the narrow \Ercm{289} resonance was not observed in this study due to an overlapping background line from carbon. We adopt the resonance parameters obtained by Bardayan \textit{et al.}~\cite{bardayan2015}. 

\noindent \textbf{E$_x$= 6.7420 (12) MeV, \Ercmb{332}, \Jpi = $3/2^-$:}
\label{6742-state}
 This state corresponds to the \Ercm{332} resonance in the \res reaction. It was observed at only three angles due to the presence of a background carbon peak in this region of the focal plane. The angular distribution for this state corresponds to $\ell=1$, indicating that it could have spin and parity \Jpi = $1/2^-$ or \Jpi = $3/2^-$. Visser \textit{et al.} \cite{visser} found that this state has \Jpi quantum numbers of $3/2^-$, later confirmed by Refs.\ \cite{nesaraja}, \cite{kahl2019}, and \cite{hall2019}. Based on this \Jpi value and on the calculated average MED, the analog level in \mirror assigned to this state is likely the one at 6.787 MeV (\Jpi = $3/2^-$).
 
\noindent \textbf{E$_x$= 6.8619 (15) MeV, \Ercmb{452}, \Jpi = $7/2^-$:}
\label{6862-state}
This state corresponds to a \Ercm{452} resonance in the \res system. It was one of the most strongly populated states in the spectra and was observed at almost all of the measured angles.  Fig.~\ref{ang_dist} shows that data for this level is best described by the curve corresponding to $\ell=3$, confirming the literature assignment of \Jpi$ = 7/2^-$ (Refs.~\cite{NNDC} and \cite{laird}). The high spin of this state creates a high angular momentum barrier, reducing its contribution to the \Fpa reaction. Based on the calculated average MED, the analog state in \mirror could be the one at 6.927 MeV. 

\noindent \textbf{E$_x$=7.0747 (9) MeV, \Ercmb{665}, \Jpi = $3/2^+$:}
\label{7076state}
This state corresponds to the \Ercm{665} resonance in the \res system. 
The resonance has a reported low spin of \Jpi = $3/2^+$ and is broad ($\Gamma_{\alpha} = 23.8$ keV) \cite{bardayan2001}. Although the width of this broad resonance should be resolvable in our experiment, the peak was in close proximity to a strong background peak, making a robust analysis challenging. This state corresponds to one of the key resonances in \Fpa at nova temperatures since it strongly interferes with other $3/2^+$ resonances, and the sign of that interference is unknown. In this experiment, the state was observed at just three angles due to the presence of a carbon peak at the focal plane. However, the measured angular distribution shown in Fig.~\ref{ang_dist} supports the literature \Jpi = $3/2^+$ assignment (e.g., \cite{NNDC}, \cite{bardayan2001}, \cite{bardayan2017}). The mirror energy level for this state has not been assigned, and the only \Jpi = $3/2^+$ in the mirror is at 7.262 MeV: a narrow state.  

\noindent \textbf{E$_x$=7.788 (9) MeV, \Ercmb{1378}, \Jpi = $\mathbf{1/2^+}$:}
\label{7790state}
The broad state predicted by Ref.~\cite{Dufour2007} corresponding to a resonance at about \Ercm{1400} was only observed at two angles in this work owing to the presence of background peaks from oxygen contamination in the target. It was also not observed by Murphy \textit{et al}~\cite{murphy}, but was at least consistent with the cross section measured by Mountford \textit{et al.}~\cite{Mountford2012}. More recently, however, Kahl \textit{et al.}~\cite{kahl2019} firmly observed an $\ell=0$ state at \Ex{7.790}, albeit with a much narrower width than that predicted in Ref.\cite{Dufour2007}. More discussion of the impact of this resonance can be found in Sec.~\ref{sec:r-matrix-analysis}.

\noindent \textbf{E$_x >$7.1 MeV:}
\label{highenergy}
Although they are not expected to have a large impact on the reaction rate at astrophysical energies, we include resonances from Mountford \textit{et al.}~\cite{Mountford2012} with the exception of their broad, inferred resonance at \Ex{1.455}, which we replace by the \Ex{7.788} state as discussed above. 

\section{R-Matrix analysis}
\label{sec:r-matrix-analysis}

In order for interference effects to significantly affect the reaction rate, the interfering resonances must have the same spin, parity, and be broad enough to overlap significantly. The R-Matrix formalism \cite{thomas-lane} provides a convenient method for investigating a reaction cross section in the presence of many broad, interfering resonances. In this formalism, the total cross section is given by \cite{desereville}:
\begin{equation}
  \label{interference}
  \sigma(E)=\frac{\pi}{k^2}\sum_{J\pi}\frac{2J+1}{(2J_0+1)(2J_1+1)}\left|U^{J\pi}(E)\right|^2,
\end{equation}
where $J$, $J_0$, and $J_1$ are the spins of the resonance, projectile, and target, respectively. $U^{J\pi}(E)$ is the collision matrix as described in Ref.~\cite{thomas-lane}. To visualize more easily the effect of resonances in the cross section, it is more useful to plot the so-called astrophysical S-factor. Using the notation above and expanding the collision matrix to resonance energies ($E_{\lambda}$) and partial widths, ($\Gamma_{\lambda}$), the astrophysical S-factor can be written as \cite{adekola-thesis}:
\begin{equation}
  \label{eq:s-factor}
 S(E)=\frac{\pi\, E\, e^{2\pi\eta} }{k^2}\sum_{J\pi}\frac{2J+1}{(2J_0+1)(2J_1+1)}\left| \sum_\lambda \frac{\pm\sqrt{\Gamma^{J\pi}_{\lambda p}(E)\,\,\Gamma^{J\pi}_{\lambda \alpha}(E)}}{E^{ J\pi}_\lambda - E - i\Gamma^{J\pi}_{\lambda}(E)}\right|^2,
\end{equation}
where $\eta$ is the Sommerfeld Parameter. The phenomenon of interference becomes clear in the sum over energy levels, $\lambda$, where for fixed \Jpi quantum numbers each term could have either a positive or negative relative sign. We will refer to the cases where the cross section increases or decreases between two resonances as constructive and destructive interference. Note, however, that these terms do \textit{not} correspond to higher and lower reaction rates, respectively, at all temperatures. Table~\ref{table-resonances} details the R-matrix input parameters used in this work to calculate the \Fpa\ cross section. 

The uncertainty in the reaction rates for the \Fpa reaction in classical nova at low temperatures (e.g., $T_9 < 80 MK$) is dominated by interference between resonances near the proton threshold and high energy resonances (e.g., 665, 1462 keV). To study the interference effects on this reaction rate, with the spin and parities found in this work for the sub-threshold resonances at -278 and -120 keV, we used the R-matrix code AZURE2 \cite{azure2}, with resonance parameters defined in Tab.\ \ref{table-resonances}. In order to remain consistent with previous studies and in light of the fact that most of the reaction rate uncertainty arises from the interference effects, we make two simplifications: (i) no uncertainties in resonance parameters are considered, and (ii) upper limit values are used as-is. The rate is likely overestimated due to the latter simplification, but the effect is minor at the temperatures of interest here. 

Although new excitation energies were determined here with comparable uncertainties to the literature, they are not used for our R-Matrix calculations. This is to ensure better consistency with the literature when comparing rates. Resonance energies are determined using the compiled resonance energies and the proton separation energy of S$_p=6410.0 (5)$~keV~\cite{AME2016}. Future work should concentrate on finalizing the resonance energies and their uncertainties to investigate their effect on the \Fpa reaction rate uncertainty.

\begin{table*}[ht]
  \centering
  \caption{List of the resonances and their properties used in the R-matrix analysis. }
  \begin{tabular}{ c c | c c c  }
    \toprule \hline
    Energy (MeV)                 & $E_R$ (keV) & {J$^{\pi}$}                       & ANC (fm$^{-1/2}$)/$\Gamma_p$ (keV)         & $\Gamma_{\alpha}$(keV)        \\ \hline
    6.1327 (23)                   & -277        & 3/2$^+$                           & 6\footnotemark[1]                          & 0.74$^a$                      \\
    6.2883 (14)                  & -122        & 1/2$^+$ + 7/2$^+$\footnotemark[7]                 & 83.5\footnotemark[4] \& 16\footnotemark[2] & 11.6 \footnotemark[5] \& 2.51 \\  
    6.416 (4) \footnotemark[3]   & 6           & 3/2$^-$                           & 4.7$\times$ 10$^{-50}$                     & 0.5                           \\
    6.439 (3) \footnotemark[3]   & 29          & 1/2$^-$                           & $<$ 3.8$\times$ 10$^{-19}$                 & 220                           \\
    6.441 (3) \footnotemark[3]   & 31          & 3/2$^+$                           & $<$ 8.4$\times$ 10$^{-18}$                 & 1.3                           \\
    6.459 (5) \footnotemark[3]   & 49          & 5/2$^-$                           & 8.4$\times$ 10$^{-14}$                     & 5.5                           \\
    6.699 (3) \footnotemark[3]   & 289         & 5/2$^+$                           & 2.4$\times$ 10$^{-5} $                     & 1.2                           \\
    6.7420 (12)                  & 332         & 3/2$^-$                           & 2.2$\times$ 10$^{-3} $ \footnotemark[5]    & 5.2 \footnotemark[5]          \\ 
    6.8619 (15)                   & 452         & 7/2$^-$                           & 1.1$\times$ 10$^{-5}$ \footnotemark[4]     & 1.2 \footnotemark[4]          \\
    7.0747 (9)                  & 665         & 3/2$^+$                           & 15.2 \footnotemark[5]                      & 23.8 \footnotemark[5]         \\
    7.170 (20) \footnotemark[6]  & 759         & 3/2$^+$                           & 1.6 (5)                                    & 2.4 (6)                       \\                        
    7.507 (11) \footnotemark[6]  & 1096        & 5/2$^+$                           & 3 (1)                                      & 54 (12)                       \\
    7.571 (34) \footnotemark[6]  & 1160        & 3/2$^+$                           & 2.3 (6)                                    & 1.9 (6)                       \\
    7.629 (22) \footnotemark[6]  & 1219        & 3/2$^-$                           & 21 (3)                                     & 0.1 (1)                       \\
    7.745 (6)  \footnotemark[6]  & 1335        & 3/2$^+$                           & 65 (8)                                     & 26 (4)                        \\    
    7.788 (9) \footnotemark[1]  & 1378        & 1/2$^+$                           & $83^{+56}_{-82}$                           & $47^{-46}_{+92}$              \\
    7.982 (13) \footnotemark[6]  & 1571        & 5/2$^+$                           & 1.7 (4)                                    & 12 (3)                        \\
    \hline\hline
    \label{table-resonances}
  \end{tabular}
  \footnotetext[1]{From Kahl \textit{et al.} 2019~\cite{kahl2019}}
  \footnotetext[2]{Upper Limit}
  \footnotetext[3]{From Hall \textit{et al.} 2020~\cite{hall2020}}
  \footnotetext[4]{From Laird \textit{et al.} 2013~\cite{laird}}
  \footnotetext[5]{From Bardayan \textit{et al.} 2015~\cite{bardayan2015}}
  \footnotetext[6]{From Mountford \textit{et al.} 2012~\cite{Mountford2012}}
  \footnotetext[7]{Maximizing spin-parity (see text for details)}
\end{table*}

Since no values have been measured for alpha decay partial widths, $\Gamma_{\alpha}$ or proton Asymptotic Normalization Coefficients (ANCs) reported for the 6.288 MeV state with \Jpi $\ge 7/2^+$, they were estimated from the mirror nucleus. To investigate the maximizing contribution of this high-spin resonance, we adopt the minimum \Jpi$=7/2^+$ spin-parity in our cross-section calculation. For the $\alpha$-particle width, the reported value  of $\Gamma_{\alpha}$ = 3.36 keV \cite{bardayan-alpha} for the analogous state in $^{19}$F at 6.330 MeV, \Jpi = 7/2$^+$ was scaled using the equation \cite{nesaraja}: 
\begin{equation}
\label{alpha-width}
 [\Gamma_{\alpha}]^{^{19}\text{Ne}} = \left[\frac{\rho}{F^2 + G^2}\right]^{^{15}\text{N} +\alpha}\left[\frac{F^2 + G^2}{\rho}\right]^{^{15}\text{O} +\alpha} [\Gamma_{\alpha}]^{^{19}\text{F}},
\end{equation}
where $\rho$ is the product of the wave number and the interaction radius, and F and G are the regular and irregular Coulomb wave functions, respectively. The proton ANC was determined using the following equation derived from %
\cite{pogrebnyak2013} and \cite{mukhamedzhanov2017}:
\begin{equation}
\label{ANC-EQ}
 \text{ANC} = \left(\frac{2 \theta^2}{W^2R}\right)^{1/2}.
\end{equation}
In this equation R is the interaction radius between a single proton and a $^{18}$F core ($R = 1.25(A_T^{1/3} + A_p^{1/3}) = 4.53$~fm), $W$ is the Whittaker function for the proton, and $\theta^2$ is the dimensionless reduced width. To evaluate the maximum contribution of the 7/2$^+$ component of the 6.288 MeV state, its ANC is assigned a value of $\theta^2$ = 1 to calculate the upper limit. The obtained values for the $\Gamma_{\alpha}$ and the ANC are shown in the corresponding column in Table \ref{table-resonances}.

\begin{figure*}
    \centering
    \includegraphics[width=0.95\textwidth]{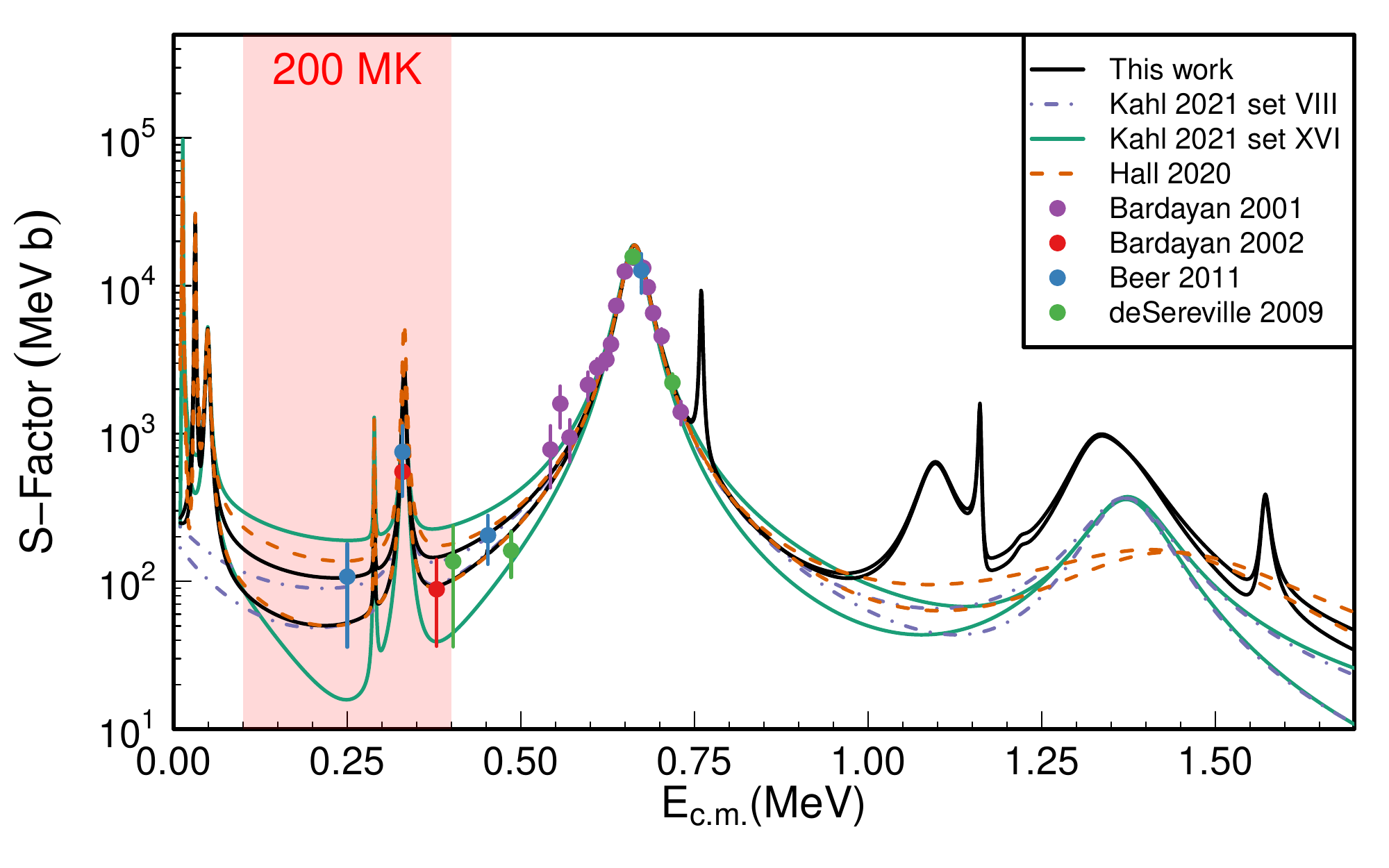}
    \caption{Astrophysical S-factor for this work (black lines), Kahl \textit{et al.} 2021 \cite{kahl2021} set VIII (purple dot-dashed lines) and set XVI (solid green lines), and Hall \textit{et al.} 2020~\cite{hall2020} (orange dashed lines). To calculate interference, relative signs for each resonance are entered into AZURE2. The signs used to obtain the maximum and minimum rates correspond to the $1/2^+$ resonances (-120, 1380 keV) and the $3/2^+$ resonances (-278, 31, 665 keV). For all rates, the maximum is obtained using relative signs (+,-) for the $1/2^+$ resonances and (-,-,+) for the  $3/2^+$ resonances. The minimum ``destructive'' rate is obtained by assigning the same sign to all resonances: (+,+) and (+,+,+).}
    \label{sfactor}
\end{figure*}

Figure \ref{sfactor} shows the astrophysical S-factors obtained in this work (black lines), the results of Hall \textit{et al.} 2020~\cite{hall2020} (orange) and the calculations presented in Kahl \textit{et al.} 2021 \cite{kahl2021} set VIII (purple) and set XVI (green lines). The S-factor is shown over a broad energy range. Although the higher-energy region does not directly affect the reaction rate at astrophysically-interesting temperatures, we include the parameters of higher-lying resonances to highlight the fact that the unknown interference terms cannot be determined trivially by measuring the high-energy cross section as one might assume from extrapolating the S-factors from the Hall/Kahl parameters.

The S-factor determined in Hall \textit{et al.} 2020 is similar at low energies to our results, with only minor differences arising from the width and location of the $\approx$1400 keV resonance. 
Here, we adopt the $E_{c.m.}$ = 1380 keV resonance parameters from Ref. \cite{kahl2019} and \cite{kahl2021} rather than the $E_{c.m.}$ = 1461 keV resonance assumed by Refs.~\cite{bardayan2017}, \cite{hall2020}, and references therein. This result is in disagreement with the theoretical prediction of a broader state at this energy \cite{Dufour2007} and subsequent inferred resonance in Ref.~\cite{Mountford2012}. However, Ref.~\cite{kahl2019} clearly observed a narrower resonance at $E_{c.m.}$ = 1380 keV in their study, confirming earlier indications from Ref.~\cite{murphy}. The proton and $\alpha$-particle partial widths are computed using the $\Gamma_p/\Gamma_{\alpha}$ branching ratio from Ref. \cite{nesaraja}.

\begin{figure*}
    \centering
    \includegraphics[width=0.95\textwidth]{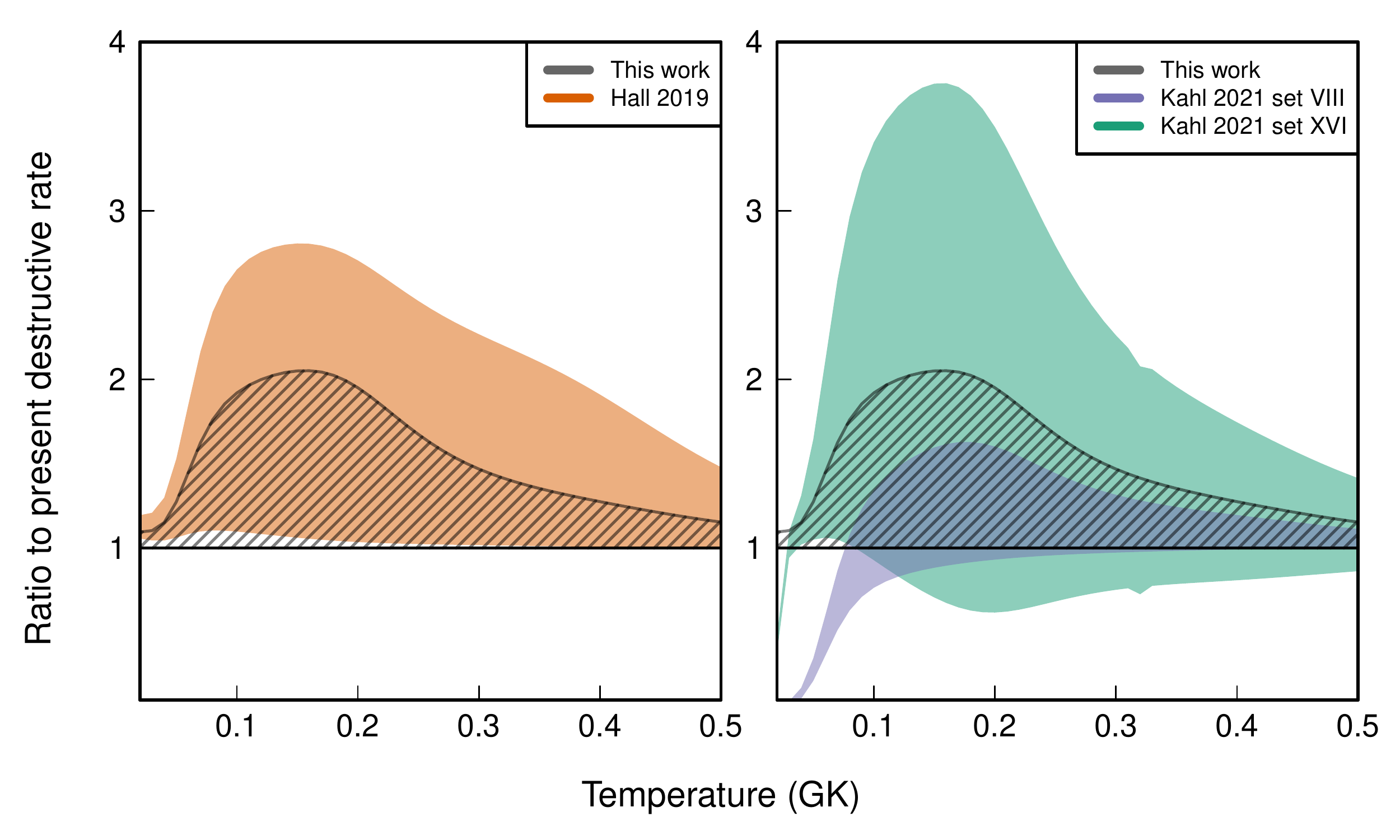}
        \caption{\Fpa reaction rates ratios between constructive interference and this work's destructive interference for Kahl \textit{et al.} \cite{kahl2021}, set XVI (green shade) and set VIII (purple shade), Hall \textit{et al.} 2020~\cite{hall2020} (orange shade), and this work (grey hatched region). The solid black line at unity represents the case were both interference possibilities are the same. The figure shows that unknown interference causes a factor of about 2 uncertainty with our present rates at 150 MK in contrast to factors of 3 and 4 in the Hall \textit{et al.} 2019 and Kahl \textit{et al.} 2020 set XVI rates, respectively.}
    \label{rates-ratios}
\end{figure*}


To more readily compare the reaction rates, Fig.~\ref{rates-ratios} displays their ratios to a common norm. We follow a similar procedure to Ref.~\cite{bardayan2015}, normalizing the different rate possibilities to our minimum reaction rate. Over the temperature region of interest, this coincides with our destructive interference case shown in Fig. \ref{sfactor}. The  left-hand panel of Fig.~\ref{rates-ratios} compares the Hall \textit{et al.} 2020 rates to the present rates, while the right-hand panel displays the ratio of the Kahl \textit{et al.} 2021~\cite{kahl2021} sets VIII and XVI rates. Our determined rates are displayed as a grey hatched region in both panels. 

Clearly, in comparison to the Hall \textit{et al.} 2020 reaction rates, the uncertainty from unknown interference is significantly reduced. At 150 MK, the uncertainty decreases from a factor of almost 3 to a factor of 2. This is because the spin-parity assignments for the resonances at \Ercm{-278} (\Jpi = $3/2^+$) and \Ercm{-120} (\Jpi = $1/2^+$) determined in this work reduce the effect of interference. At higher temperatures, the uncertainty is reduced for two reasons: (i) including higher-energy resonances reduces the impact of interference effects, and (ii) the 1.4 MeV resonance adopted here is narrower than the one assumed by Ref.~\cite{hall2020}, slightly reducing the span of cross sections that interference produces.

The reaction rates presented in Kahl \textit{et al.} 2021~\cite{kahl2021} are arguably a much more exhaustive study that considered the combinations of resonances allowed by the literature. Their Set XVI was reported to produce the maximum constructive interference effect. The sub-threshold spin-parity results of our work now rule out that possibility, thus significantly reducing the range of reaction rates by almost a factor of 2 at 150 MK. The large difference arises almost entirely because of the order of the sub-threshold $1/2^+$ and $3/2^+$ resonances, which is reversed between their calculation and our experimental results. Conversely, the interference effects computed using our resonance parameters are larger than those predicted using the Kahl \textit{et al.} set VIII parameters. Our reaction rate is considerably faster than theirs below 100 MK. Although their set VIII parameters most closely resemble our results, both of these effects are explained by the additional low energy resonances included here at \Ex{6.416 - 6.459}.

\section{Conclusions}
\label{sec:conclusions}

A \reac neutron pick up reaction at 21 MeV was used to populated \Ne excited states relevant for the \Fpa reaction at astrophysical energies. It was carried out at the Triangle Universities Nuclear Laboratory (TUNL) using the Enge split-pole magnetic spectrograph. The experiment employed targets of $^{20}$Ne implanted on carbon foils, and a focal plane detector was used to collect the reaction products. A total of 29 \Ne excited states were populated and their energy determined. Angular distributions were used to extract the spin and parity of \Ne excited states in the astrophysical energy range (6 - 7 MeV). Crucially, the state at 6.133 MeV corresponding to the \Ercm{-278} sub-threshold resonance was found to have a \Jpi value of $3/2^+$, and the 6.288 MeV state (\Ercm{-120} sub-threshold resonance) was observed as an unresolved doublet with a combination of \Jpi quantum numbers:  $1/2^+$ and $\ge 7/2^+$.

Previous studies of this reaction (Refs.~\cite{hall2020,kahl2021} being the most recent) showed that ambiguous spin-parity assignments to states around the proton threshold lead to large uncertainties in the \Fpa cross section due to unknown interference. R-matrix calculations were performed to investigate the impact of our spin-parity assignments on these effects. The cross sections and reaction rates derived here are compared with those of Hall \textit{et al.} 2020~\cite{hall2020} and Kahl \textit{et al.} 2021~\cite{kahl2021} (sets VIII and XVI). We find that our spin-parity assignments significantly reduce the uncertainty in the \Fpa cross section arising from interference effects by up to a factor of two. 

In light of our findings, the remaining questions regarding the \Fpa reaction are becoming clearer. A full re-analysis and accounting for mirror states between \Ne and \mirror should be performed to understand the issue of unmatched states around the proton threshold. A statistical treatment of the remaining ambiguities and uncertainties in the nuclear structure of \Ne would then be possible. Furthermore, complete nucleosynthesis studies may elucidate where, or if, further experimental information is required.

\section{Acknowledgements}
We would like to thank Nicolas de S\'er\'eville and Fa\"irouz Hammache
for their very insightful discussion on some aspects of this work.
This material is based upon work supported by the U.S. Department of
Energy, Office of Science, Office of Nuclear Physics, under Award
Numbers DE-SC0017799 and under Contract No. DE-FG02-97ER41041. This
paper is also based upon work from the ‘ChETEC’ COST Action (CA16117),
supported by COST (European Cooperation in Science and Technology)


\end{document}